\documentstyle[preprint,aps,epsf,floats]{revtex}

\newcommand{\nn}{\nonumber}

\begin{document}
\setlength\baselineskip{20pt}

\preprint{\tighten \vbox{\hbox{CALT-68-2242} \hbox{UCSD/PTH 99-14}
}}

\title{Conformal Invariance for Non-Relativistic Field Theory
}
\author{Thomas Mehen$^a$, Iain W. Stewart$^{b}$, and Mark B. Wise$^a$\\[15pt]}
\address{\tighten $^a$California Institute of Technology, Pasadena, CA
91125\\[5pt] $^b$Department of Physics, University of California at San
Diego,\\[2pt] 9500 Gilman Drive, La Jolla, CA 92099\\[15pt]}

\maketitle

{\tighten
\begin{abstract}

Momentum space Ward identities are derived for the amputated n-point Green's
functions in $3+1$ dimensional non-relativistic conformal field theory.  For
$n=4$ and $6$ the implications for scattering amplitudes (i.e. on-shell
amputated Green's functions) are considered.  Any scale invariant 2-to-2
scattering amplitude is also conformally invariant. However, conformal
invariance imposes constraints on off-shell Green's functions and the three
particle scattering amplitude which are not automatically satisfied if they are
scale invariant. As an explicit example of a conformally invariant theory we
consider non-relativistic particles in the infinite scattering length limit.

\end{abstract}
}
\vspace{0.7in}

\newpage

Poincar\'{e} invariant theories that are scale invariant usually have a larger
symmetry group called the conformal group\footnote{\tighten Exceptions are
known to exist, however, these theories suffer from pathologies, such as
non-unitarity. A detailed discussion of scale and conformal invariance in
relativistic theories can be found in Ref.\cite{joej}.}. A similar phenomena
happens for 3+1 dimensional non-relativistic systems. These are invariant under
the extended Galilean group, which consists of 10 generators: translations (4),
rotations (3), and Galilean boosts (3). The largest space-time symmetry group
of the free Schr\"{o}dinger equation is called the Schr\"{o}dinger or
non-relativistic conformal group \cite{nrconf}. This group has two additional
generators corresponding to a scale transformation, and a one-dimensional
special conformal transformation, sometimes called an ``expansion''. The
infinitesimal Galilean boost, scale and conformal transformations are
\begin{eqnarray} \label{infs}
 \begin{tabular}{llllll}
 & boosts:\qquad & & $\vec x\,' = \vec x + \vec v\, t$ \,,  & &  $t'=t$ \,,\\
 & scale:\qquad  & & $\vec x\,' = \vec x + s \vec x$ \,,  & & $t'=t + 2 s t$
   \,, \\
 & conformal:\qquad\qquad & & $\vec x\,' = \vec x - c t \vec x$ \,, & & $t'
   = t - c t^2$ \,, \end{tabular}
\end{eqnarray}
where $\vec v$, $s$ and $c$ are the corresponding infinitesimal parameters.
(The finite scale transformation is $\vec x\,'=e^s\,\vec x$, $t'=e^{2s}\,t$, and
the finite conformal transformation is $\vec x\,'=\vec x/(1+ c t)$,
$1/t'=1/t+c$.)

In this letter we explore the implications of non-relativistic conformal invariance
for $3+1$ dimensional physical systems. In relativistic theories, conformal
invariance can be used to constrain the functional form of n-point correlation
functions\cite{joej}, however, on-shell scattering amplitudes are typically
ill-defined because of infrared divergences associated with massless particles.
In non-relativistic theories scattering amplitudes are well defined even in the
conformal limit. We show how conformal invariance can be used to gain
information about scattering amplitudes by deriving Ward identities for the
amputated momentum space Green's functions.  While the off-shell Green's
functions can be changed by field redefinitions, the scattering amplitudes
(on-shell Green's functions) are physical quantities and are therefore
unchanged. We find that any 2-to-2 (identical particle) scattering amplitude that
satisfies the scale Ward identity automatically satisfies the conformal Ward
identity.  However, this is not the case for the corresponding off-shell Green's
function or for the 3-to-3 scattering amplitude.  We construct a field theory that
has a four point function which obeys the scale and conformal Ward identities
and conjecture that the higher point functions in this theory also obey these
Ward identities.  On-shell it gives S-wave scattering with an infinite scattering
length.

For the interaction of two nucleons, the scattering lengths in the $^1S_0$ and
$^3S_1$ channels are large ($a^{(^1S_0)} = -23.7 \,{\rm fm}$ and $a^{(^3S_1)} =
5.4 \,{\rm fm}$) compared to the typical length scales in nuclear physics. In
the limit that these scattering lengths go to infinity (and higher terms in the
effective range expansion are neglected) we show that the four point Green's
function obeys the scale and conformal Ward identities. Thus, two body nuclear
systems at low energies are approximately scale and conformal invariant. It is
likely that in some spin-isospin channels the higher point functions will also
obey these Ward identities. Whether this conformal invariance can lead to new
predictions for many body nuclear physics is presently unclear, but seems
worthy of further study.

The action for a free non-relativistic field $N(\vec x,t)$ is
\begin{eqnarray} \label{S0}
  S_0 = \int dt d^3x \ N^\dagger \Big( i\partial_t +\frac{\nabla^2}{2M} \Big)
  N\,,
\end{eqnarray}
where $M$ is mass of the particle corresponding to the field $N$. Under an
infinitesimal Galilean transformation $N'(\vec x',t')= (1+ i M\, \vec v\cdot
\vec x) N(\vec x,t)$ or equivalently
\begin{eqnarray} \label{trng}
 \delta_g N(\vec x,t) = N'(\vec x,t)-N(\vec x,t) = {\rm D}_g\:
  N(\vec x,t) = \vec v \cdot \Big( i M\, \vec x - t \vec\nabla \Big) N(\vec x,t)
  \,.
\end{eqnarray}
The action  in Eq.~(\ref{S0}) is invariant under the infinitesimal scale
transformation in Eq.~(\ref{infs}) with $N'(\vec x',t')=(1-3 s/2) N(\vec x,t)$
or equivalently
\begin{eqnarray} \label{trns}
  \delta_s N(\vec x,t) =  {\rm D}_s\: N(\vec x,t) = -s \Big( \frac{3}{2} +
  \vec x \cdot \vec\nabla + 2 t \partial_t \Big) N(\vec x,t) \,,
\end{eqnarray}
and under the infinitesimal conformal transformation provided $N'(\vec x',t') =
(1+3\, c\, t/2 -i M\, c\, \vec x\,^2 /2 ) N(\vec x,t)$ or equivalently
\begin{eqnarray} \label{trnc}
  \delta_c N(\vec x,t) =  {\rm D}_c\: N(\vec x,t) = c \Big( \frac{3t}{2}-
  \frac{i M  \vec x\,^2}{2} + t \vec x \cdot \vec \nabla + t^2 \partial_t \Big)
  N(\vec x,t)  \,.
\end{eqnarray}

Now consider adding interactions that preserve these invariances (an explicit
example will be considered later).  The position space Green's functions for
the interacting theory, $G^{(2n)}(\vec x_i,t_i)=G^{(2n)}(\vec x_1,t_1;\ldots ;
\vec x_{2n}, t_{2n})$, are defined by\footnote{\tighten In non-relativistic
theories particle number is conserved so there must be the same number of $N$'s
as $N^\dagger$'s.}
\begin{eqnarray}
  G^{(2n)}(\vec x_i,t_i) = \langle\: \Omega |\, T\Big\{\: N(\vec x_1,t_1)\cdots
  N(\vec x_n,t_n) N^\dagger(\vec x_{n+1},t_{n+1})\cdots
  N^\dagger(\vec x_{2n},t_{2n})\:\Big\} | \Omega \rangle \,,
\end{eqnarray}
where $|\Omega\rangle$ is the vacuum of the interacting theory and is assumed
to be invariant under the Schr\"{o}dinger group. Under the infinitesimal
transformations in Eqs.~(\ref{trng}-\ref{trnc})
\begin{eqnarray} \label{varG}
  \delta_{(g,s,c)} G^{(2n)}(\vec x_i,t_i) &=& \langle \Omega |\: T \Big\{\:
  \delta_{(g,s,c)} N(\vec x_1,t_1) N(\vec x_2,t_2)\cdots N^\dagger
  (\vec x_{2n},t_{2n})\:\Big\} | \Omega \rangle + \ldots \nn \\
  && + \langle \Omega | \:T \Big\{\:  N(\vec x_1,t_1) \cdots
  N^\dagger(\vec x_{2n-1},t_{2n-1}) \delta_{(g,s,c)}
  N^\dagger(\vec x_{2n},t_{2n})\: \Big\} | \Omega \rangle  \nn\\
  &=& \Big[ \sum_{k=1}^{n} {\rm D}^k_{(g,s,c)} + \sum_{k=n+1}^{2n}
  {\rm D}^{k\,\dagger}_{(g,s,c)}  \Big] \langle\: \Omega |\,
  T\Big\{\: N(\vec x_1,t_1)\cdots N^\dagger(\vec x_{2n},t_{2n})\:\Big\}
  | \Omega \rangle \,,
\end{eqnarray}
where ${\rm D}_{(g,s,c)}^k$ is the differential operator for coordinates $(\vec
x_k,t_k)$.  Invariance under Galilean boosts, scale, and conformal symmetry
implies that
\begin{eqnarray} \label{spvar}
  \delta_{(g,s,c)}\: G^{(2n)}(\vec x_i,t_i) =  0 \,.
\end{eqnarray}

The momentum space Green's functions $G^{(2n)}(\vec p_i,E_i)= G^{(2n)}(\vec
p_1,E_1;\ldots; \vec p_{2n},E_{2n})$ are the Fourier transform of the  position
space Green's functions
\begin{eqnarray} \label{ft}
  && G^{(2n)}(\vec x_1,t_1;\ldots;\vec x_{2n},t_{2n}) = \bigg[ \prod_{k=1}^{2n}
  \int {dE_{k}d^3p_{k} \over (2\pi)^4} e^{-i \eta_k (E_k t_k- \vec p_k\cdot
  \vec x_k)} \bigg] \nn\\
  &&\qquad \times\ (2\pi)^4\: \delta\Big(\sum_{k=1}^{2n} \eta_k E_k\Big)\:
  \delta^{(3)}\Big(\sum_{k=1}^{2n} \eta_k \vec p_k\Big)\:  G^{(2n)}(E_1,
  \vec p_1;\ldots; E_{2n},\vec p_{2n}) \,,
\end{eqnarray}
where $\eta_j$ is $1$ for incoming particles (subscripts $1,\ldots,n$) and $-1$
for outgoing particles (subscripts $n+1,\ldots,2n$). The delta functions in
Eq.~(\ref{ft}) arise due to translational invariance. Using Eq.~(\ref{spvar})
with $\vec x_{2n}=0$ and $t_{2n}=0$ it is straightforward to show that
invariance under Galilean boosts, scale transformations, and conformal
transformations implies the Ward identities
\begin{eqnarray} \label{wardG}
   {\cal D}_{(g,s,c)}\: G^{(2n)}(E_1,\vec p_1;\ldots;E_{2n},\vec p_{2n}) =0 \,,
\end{eqnarray}
where
\begin{eqnarray}
  {\cal D}_{g} &=& \sum_{j=1}^{2n-1} \Big( M \vec\nabla_{p_j} +\vec p_j
  {\partial \over \partial{E_j}} \Big) \,, \\
  {\cal D}_{s} &=& 7 n - 5 + \sum_{j=1}^{2n-1} \Big( \vec p_j \cdot
  \vec \nabla_{p_j} + 2 E_j\, {\partial \over \partial{E_j}} \Big) \,, \nn \\
  {\cal D}_{c} &=& \sum_{j=1}^{2n-1} \eta_j\, \Big( \frac72\, {\partial \over
  \partial{E_j}} + \frac{M}{2} \vec \nabla^2_{p_j} + E_j\, {\partial^2 \over
  \partial{E_j}^2} +\vec p_j \cdot \vec \nabla_{p_j} {\partial \over
  \partial{E_j}} \Big) \,. \nn
\end{eqnarray}
In deriving Eq.~(\ref{wardG}) we have integrated over the delta functions in
Eq.~(\ref{ft}) so that
\begin{eqnarray}  \label{trans}
  E_{2n}= \sum_{j=1}^{2n-1} \eta_j E_j \,,\qquad
  {\vec p_{2n}} = \sum_{j=1}^{2n-1} \eta_j {\vec p}_j \,.
\end{eqnarray}
The S-matrix elements are related to the amputated Green's functions ${\cal
A}^{(2n)}(\vec p_i,E_i)={\cal A}^{(2n)}(\vec p_1,E_1;\ldots; \vec
p_{2n},E_{2n})$ defined by\footnote{\tighten Neglecting relativistic corrections
to $S_0$, Eq.~(\ref{defA}) is exact because adding interactions to
Eq.~(\ref{S0}) does not effect the two point function since there is no pair
creation in the non-relativistic theory.}
\begin{eqnarray} \label{defA}
 {\cal A}^{(2n)}(E_i,\vec p_i) = \bigg[ \prod_{j=1}^{2n} \Big(E_j-\frac{\vec
  p_j\,^2}{2M} \Big)\:\bigg]\ G^{(2n)}_{con.}(E_i,\vec p_i) \,,
\end{eqnarray}
where $E_{2n}$ and $\vec p_{2n}$ are given by Eq.~(\ref{trans}).
$G^{(2n)}_{con.}$ is the connected part of $G^{(2n)}$ and also satisfies
Eq.~(\ref{wardG}). Applying the Galilean boost and scale Ward identities in
Eq.~(\ref{wardG}) to Eq.~(\ref{defA}) gives
\begin{eqnarray} \label{wardA}
  \tilde {\cal D}_{(g,s)} \: {\cal A}^{(2n)}(E_i,\vec p_i) = 0 \,,
\end{eqnarray}
where $\tilde {\cal D}_g = {\cal D}_g$ and
\begin{eqnarray}
  \tilde {\cal D}_{s} &=& 3n-5 + \sum_{j=1}^{2n-1}
  \Big( \vec p_j \cdot \vec\nabla_{p_j} + 2 E_j\, {\partial \over
  \partial{E_j}} \Big) \,.
\end{eqnarray}
Applying the conformal Ward identity in Eq.~(\ref{wardG}) to Eq.~(\ref{defA})
gives
\begin{eqnarray}
  \tilde {\cal D}_{c} {\cal A}^{(2n)} + {1 \over \sum_j \eta_j E_j - (\sum_j
  \eta_j \vec p_j)^2/(2M) } \bigg[\frac{1}{M} \Big(\sum_j \eta_j \vec p_j
  \Big)\cdot {\cal D}_g - \tilde {\cal D}_s \bigg] {\cal A}^{(2n)} =0 \,,
\end{eqnarray}
where
\begin{eqnarray}
  \tilde {\cal D}_{c} &=& \sum_{j=1}^{2n-1} \eta_j\, \Big( \frac32\,
  {\partial \over \partial{E_j}}
  + \frac{M}{2}\, \vec \nabla^2_{p_j} + E_j\, {\partial^2 \over \partial{E_j}^2}
  + \vec p_j \cdot {\vec \nabla}_{p_j}\: {\partial \over \partial{E_j}} \Big) \,.
\end{eqnarray}
Therefore, amputated Green's functions satisfying Eq.~(\ref{wardA}) also
satisfy
\begin{eqnarray} \label{wardA2}
  \tilde {\cal D}_c \: {\cal A}^{(2n)}= 0 \,.
\end{eqnarray}

\begin{figure}[!t]
  \centerline{\epsfxsize=18.5truecm \epsfbox{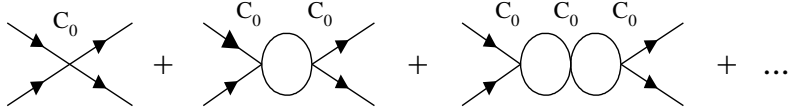}  }
 {\tighten
\caption[1]{Terms contributing to ${\cal A}^{(4)}$ from the interaction in
Eq.~(\ref{S1}).} \label{C0b} }
\end{figure}

The leading term in the effective field theory for non-relativistic
nucleon-nucleon scattering corresponds to a scale invariant theory in the limit
that the S-wave scattering lengths go to infinity (see for e.g.
Ref.~\cite{msw}). As we will see below, this limit corresponds to a fixed point
of the renormalization group.  Since in nature the S-wave scattering lengths are
very large, it is the unusual scaling of operators at this non-trivial fixed
point\cite{Weinberg} that controls their importance in this effective field
theory \cite{ksw,birse}. Motivated by this we add to Eq.~(\ref{S0}) the
interaction
\begin{eqnarray} \label{S1}
  S_1 &=& -\int dt d^3x\ C_0\: (N^T P N)^\dagger (N^T P N) \,,
\end{eqnarray}
where $N$ is now a two component spin-$1/2$ fermion field and $P=i\sigma_2/2$.
Higher body non-derivative interaction terms are forbidden by Fermi statistics.
The interaction in Eq.~(\ref{S1}) only mediates spin singlet S-wave $NN$
scattering. The $NN$ scattering amplitude arises from the sum of bubble Feynman
diagrams shown in Fig.~\ref{C0b}.
The loop integration associated with a bubble has a linear ultraviolet
divergence and consequently the values of the coefficients $C_0$ depend on the
subtraction scheme adopted. In minimal subtraction, if $p\gg 1/a$ where $p$ is
the center of mass momentum and $a$ is the scattering length, then successive
terms in the perturbative series represented by Fig.~\ref{C0b} get larger and
larger. Subtraction schemes have been introduced where each diagram in
Fig.~\ref{C0b} is of the same order as the sum. One such scheme is PDS
\cite{ksw}, which subtracts not only poles at $D=4$, but also the poles at
$D=3$ (which correspond to linear divergences). Another such scheme is the OS
momentum subtraction scheme \cite{Weinberg,OS}.  In these schemes the
coefficients are subtraction point dependent, $C_0 \equiv C_0(\mu)$. Calculating
the bubble sum in PDS or OS gives
\begin{eqnarray} \label{A4mu}
   {\cal A}^{(4)} = {- C_0(\mu) \over 1 + M C_0(\mu) \Big[\: 2\mu -
   \sqrt{-4M(E_1+E_2) +(\vec p_1 + \vec p_2)^2 - i\epsilon}\ \Big]/(8\pi) }\,,
\end{eqnarray}
where
\begin{eqnarray}
   C_0(\mu) = -\frac{4\pi}{M} \frac1{\mu -1/a} \,.
\end{eqnarray}
Note that Eq.~(\ref{A4mu}) holds in any frame and we have not imposed the
condition that the external particles be on-shell. It is easy to see that the
limit $a \rightarrow \pm\infty$ corresponds to a nontrivial ultraviolet fixed
point in this scheme. If we define a rescaled coupling ${\hat C_0} \equiv M
\mu\, C_0(\mu)/(4 \pi)$, then
\begin{eqnarray}
   \mu {d \over d \mu} {\hat C_0}(\mu) =  {\hat C_0}(\mu)\Big[ 1+
   {\hat C_0}(\mu) \Big]\,.
\end{eqnarray}
The limit $a \rightarrow \pm\infty$ corresponds to the fixed point ${\hat C_0}
= -1$. At a fixed point one expects the theory to be scale invariant. In fact,
it can be easily verified that in the $a \rightarrow \pm\infty$ limit
\begin{eqnarray} \label{A41}
 {\cal A}^{(4)} = { 8 \pi \over M}{1 \over \sqrt{-4M(E_1+E_2) +(\vec p_1 +
  \vec p_2)^2 - i\epsilon}}
\end{eqnarray}
satisfies both the scale and conformal Ward identities in Eqs.~(\ref{wardA})
and (\ref{wardA2}). In the case of ${\cal A}^{(4)}$ the conformal Ward identity
gives non-trivial information about the off-shell amplitude.  For instance the
amplitude
\begin{eqnarray} \label{A42}
 {\cal A}^{(4)} = { 8 \pi \over M}{1 \over \sqrt{-(\vec p_1 - \vec p_3)^2-
 (\vec p_2-\vec p_3)^2 -i\epsilon }}
\end{eqnarray}
is scale and Galilean invariant but not conformally invariant.  The expressions
for ${\cal A}^{(4)}$ in Eqs.~(\ref{A41}) and (\ref{A42}) agree on-shell, where
$E_i=\vec p_i\,^2/(2M)$.

The interaction in Eq.~(\ref{S1}) also induces non-trivial amputated Greens
functions, ${\cal A}^{(2n)}$, for $n>2$. (For $n=3$ see Fig.~2.) It is believed
that non-perturbatively the higher point functions are finite and we speculate
that with $C_0$ at its critical fixed point the action $S_0+S_1$ defines a
non-relativistic conformal field theory.
\begin{figure}[!t]
  \centerline{\epsfxsize=20.5truecm \epsfbox{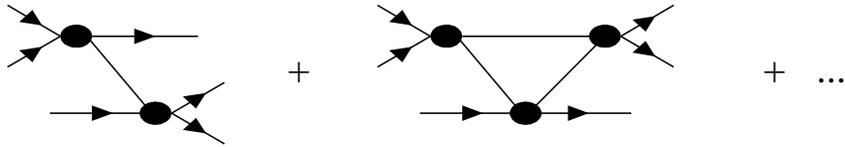}  }
 {\tighten \caption[1]{Terms contributing to ${\cal A}^{(6)}$ from the interaction in
Eq.~(\ref{S1}).  The filled circle denotes the sum of diagrams in Fig.~1.}}
\end{figure}

We will now derive scale and conformal Ward identities for the on-shell
amplitudes since these are the physical quantities of interest. Consider the four
point function for a scalar field\footnote{\tighten Eqs.~(\ref{A4start}) through
(\ref{a4gal}) are valid for fermions, but when imposing rotational invariance we
will assume the particles are scalars. For fermions ${\cal A}^{(4)}$ has spin
singlet and spin triplet parts, and the expressions in Eq.~(\ref{a4Rot}) are
valid for the spin singlet component.}. After imposing translation invariance it
is a function of 12 variables
\begin{eqnarray} \label{A4start}
   {\cal A}^{(4)}(\vec p_1, \vec p_2, \vec p_3, E_1, E_2, E_3) \,.
\end{eqnarray}
The Ward identity $\tilde {\cal D}_g {\cal A}^{(4)}=0$ is solved by the
function ${\cal A}^{(4)}(\vec p_A,\vec p_B, U_1,U_2,U_3)$ where
\begin{eqnarray}
  \vec p_A =\vec p_1-\vec p_3 \,,\qquad \vec p_B=\vec p_2-\vec p_3\,, \qquad
  U_i = M E_i -{\vec p_i\,^2 \over 2}  \,.
\end{eqnarray}
Therefore, using the Galilean boost invariance gives three constraints on
${\cal A}^{(4)}$ leaving 9 variables. For this function the scale and conformal
identities are
\begin{eqnarray} \label{a4gal}
 \tilde {\cal D}_s &=& 1 + \vec p_A \cdot \vec \nabla_{p_A} + \vec p_B \cdot
 \vec \nabla_{p_B} + 2 \sum_{j=1}^3 U_j {\partial \over \partial U_j} \,, \nn\\
 \tilde {\cal D}_c &=& M\bigg( - \vec \nabla_{p_A} \cdot
 \vec \nabla_{p_B} +  \sum_{j=1}^3 \eta_j\: U_j {\partial^2 \over \partial
 U_j^2} \bigg) \,.
\end{eqnarray}
Three more constraints are given by rotation invariance leaving a function of 6
variables, ${\cal A}^{(4)}(x,y,\gamma,U_1,U_2,U_3)$, where
\begin{eqnarray}
  x = \vec p_A\,^2 \,,\qquad y= \vec p_B\,^2 \,, \qquad
  \gamma = \vec p_A \cdot \vec p_B \,.
\end{eqnarray}
In terms of these variables we have
\begin{eqnarray} \label{a4Rot}
 \tilde {\cal D}_s &=& 1 + 2 x { \partial \over \partial x} + 2 y {\partial
 \over \partial y} + 2\gamma {\partial \over \partial \gamma} + 2 \sum_{j=1}^3
 U_j {\partial \over \partial U_j} \,, \nn \\
 \tilde {\cal D}_c &=& -M \bigg( {\partial \over \partial \gamma} \tilde
 {\cal D}_s + 4\gamma { \partial^2 \over \partial x \partial y} -\gamma
 {\partial^2 \over \partial \gamma^2} - 2 \sum_{j=1}^3 U_j {\partial^2 \over
 \partial\gamma \partial U_j} - \sum_{j=1}^3 \eta_j\:U_j {\partial^2 \over
 \partial U_j^2} \bigg)  \,.
\end{eqnarray}
On-shell the four point function has an additional four constraints
$U_1=U_2=U_3=0$ and $\gamma=0$, where the last condition follows because
$U_4=U_1+U_2-U_3-\gamma=0$. The operator $\tilde {\cal D}_s$ can be defined
consistently on-shell since all derivatives with respect to $U_{1,2,3}$ and
$\gamma$ are multiplied by coefficients which vanish in the on-shell limit. In
taking the on-shell limit we are assuming that derivatives of ${\cal A}^{(4)}$
with respect to the off-shell parameters are not singular. This is true of the
explicit example in Eq.~(\ref{A41}) as long as the momentum of the nucleons in
the center of mass frame is nonzero.  Finally, from Eq.~(\ref{a4Rot}) we see
that on-shell a scale invariant ${\cal A}^{(4)}$ is automatically conformally
invariant.

Solving $\tilde {\cal D}_s {\cal A}^{(4)}=0$, the most general scattering
amplitude consistent with Schr\"{o}dinger group invariance is
\begin{eqnarray} \label{A4on}
  A^{(4)}_{os} = \frac{1}{\sqrt{x+y}}\: F\Big( \frac{y-x}{y+x} \Big) =
  \frac{1}{2p}\: F(\cos\theta) \,,
\end{eqnarray}
where $F$ is an arbitrary function, and $\theta$ is the scattering angle in the
center of mass frame. Conformal invariance does not restrict the
angular dependence of the scattering amplitude. Additional physical criteria
can be used to provide further constraints. The condition that the S-wave
scattering length goes to infinity corresponds to a fine tuning that produces a
bound state at threshold.  Assuming that this is the only fine tuning and that
the interactions are short range the threshold behavior of the phase shift in the
$\ell$th partial wave is $\delta_\ell \sim p^{2\ell+1}$ for $\ell>0$. It is
easy to see that the only partial wave obtained from Eq.~(\ref{A4on}) with
acceptable threshold behavior is the S-wave, so $F$ can be replaced by a
constant. In the limit $a\to \infty$ the interaction in Eq.~(\ref{S1}) provides
an explicit example of a scale invariant theory which has this behavior.

In the case of the 3-to-3 scattering amplitude, conformal invariance will
provide a new constraint independent from that of scale invariance. We proceed
exactly as in the case of the 2-to-2 scattering amplitude. After imposing
energy and momentum conservation the 6 point function has 20 coordinates
\begin{eqnarray}
   {\cal A}^{(6)}(\vec p_1,\ldots \vec p_5, E_1,\ldots, E_5) \,.
\end{eqnarray}
Using the Galilean boost invariance leaves 17 coordinates
\begin{eqnarray}
 {\cal A}^{(6)}(\vec p,\vec k,\vec p\,',\vec k\,', U_1,\ldots,U_5)\,,
\end{eqnarray}
where $U_i = M E_i - {\vec p_i\,}^2/2$ and
\begin{eqnarray}
  \vec p &=& \frac{2\vec p_3-\vec p_2-\vec p_1}{3}, \qquad\qquad\qquad\qquad\
  \vec k=\vec p_2-\vec p_1, \\
  \vec p\,'&=& \frac{2(\vec p_1+ \vec p_2 + \vec p_3)}{3}-\vec p_4-\vec p_5,
  \qquad\quad \vec k\,' = \vec p_5-\vec p_4 \,. \nn
\end{eqnarray}
In terms of these variables
\begin{eqnarray}
 \tilde {\cal D}_s &=& 4 + \vec p\cdot \vec\nabla_{p} + \vec k \cdot
 \vec\nabla_{k} + \vec p\,' \cdot \vec\nabla_{p'} + \vec k\,' \cdot
 \vec\nabla_{k'} + 2 \sum_{j=1}^5 \: U_j\: {\partial \over \partial U_j} \nn\\
 \tilde {\cal D}_c &=& \frac{M}{3} \Big[ \vec\nabla_{p}^2 + 3\vec\nabla_{k}^2
 - \vec\nabla_{p'}^2 - 3\vec\nabla_{k'}^2 \Big] + M \sum_{j=1}^5 \eta_j\:
 U_j\: {\partial^2 \over \partial U_j^2}  \,.
\end{eqnarray}
Next consider imposing rotational invariance. For simplicity we specialize to
the case of a scalar field. Rotational invariance implies that ${\cal A}^{(6)}$
should be a function of 14 variables. We have chosen
\begin{eqnarray} \label{A6r}
 {\cal A}^{(6)}(z_1,\ldots,z_8, \gamma, U_1,\ldots,U_5)\,,
\end{eqnarray}
where
\begin{eqnarray}
 && z_1=\vec p\,^2\,,\qquad\quad z_2=\vec k\,^2 \,,\qquad\quad
 z_3=\vec p\,'\,^2 \,,\qquad\quad z_4=\vec p\cdot \vec k\,,\\
 && z_5=\vec p\cdot \vec p\,'\,,\qquad z_6=\vec p\cdot \vec k\,'\,,
 \qquad z_7=\vec k\cdot \vec p\,' \,,\qquad z_8=\vec p\,'\cdot \vec k\,' \,,\nn\\
 && \gamma = \vec k\,^2 -\vec k\,{'\,^2}+3\vec p\,\:^2 -3 \vec p\,\,'\,^2 \,.\nn
\end{eqnarray}
The coordinates $U_i$ and $\gamma$ vanish on-shell since $U_6=\sum_{j=1}^5
\eta_j\, U_j +\gamma/4$. For the function in Eq.~(\ref{A6r}) the scale and
conformal derivatives are
\begin{eqnarray}  \label{wardC3}
 \tilde {\cal D}_s &=& 4
   + 2 \sum_{j=1}^8 \: z_j\: {\partial \over \partial z_j}+\ldots \,, \nn \\
 \tilde {\cal D}_c &=&2 M\bigg( {\partial \over \partial z_1} +3 {\partial \over
 \partial z_2} -{\partial \over \partial z_3} \bigg) +\frac{M}{3} \sum_{j,k=1}^8
 A_{jk}{\partial^2 \over \partial z_j\: \partial z_k } + 4 M {\partial \over
 \partial \gamma} {\tilde D_s}+ \ldots \,.
\end{eqnarray}
The ellipses are terms with factors of $U_i$ or $\gamma$ and therefore vanish
on-shell,
\begin{eqnarray}
  A_{jk} = \left( \begin{array}{cccccccc}
   4 z_1 & 0 & 0 & 2 z_4 & 2 z_5 & 2 z_6 & 0 & 0 \\
   0 & 12 z_2 & 0 & 6 z_4 & 0 & 0 & 6 z_7 & 0 \\
   0 & 0 & -4 z_3 & 0 & -2 z_5 & 0 & -2 z_7 & -2 z_8 \\
   2 z_4 & 6 z_4 & 0 & 3 z_1+ z_2 & z_7 & z_9 & 3 z_5 & 0 \\
   2 z_5 & 0 & -2 z_5 & z_7 & -z_1+z_3 & z_8 & -z_4 & -z_6 \\
   2 z_6 & 0 & 0 & z_9 & z_8 & z_2-3 z_3 & 0 & -3 z_5 \\
   0 & 6 z_7 & -2 z_7 & 3 z_5 & -z_4 & 0 & -z_2+3 z_3 & -z_9 \\
   0 & 0 & -2 z_8 & 0 & -z_6 & -3 z_5 & -z_9 & -3 z_1-z_2 \end{array} \right) \,,
\end{eqnarray}
and $z_9 = \vec k \cdot \vec k\,'$. It is possible to express $z_9$ in terms of
$z_1,\ldots,z_8$. For scale invariant amputated Green's functions the conformal
operator can be defined on-shell because terms that involve derivatives with
respect to the off-shell parameters ($U_i$ and $\gamma$) have coefficients
which vanish on-shell.

Even after demanding scale invariance the conformal Ward identity still imposes
a nontrivial constraint on the amplitude. It is easy to find examples of boost
and scale invariant functions which do not satisfy $\tilde {\cal D}_c\, {\cal
A}^{(6)}=0$. Due to the complexity of Eq.~(\ref{wardC3}) we have not attempted
to find its general solution.

The effective field theory for the strong interactions of nucleons is more
complicated than the toy model given by $S_0+S_1$, because nucleons have
isospin degrees of freedom. The inclusion of internal degrees of freedom does
not change the Ward identities that correlations must satisfy to be
Schr\"{o}dinger invariant. However, isospin allows additional contact
interactions to exist. There are two four-nucleon operators ($^1S_0$ and
$^3S_1$) and one six-nucleon operator that can be formed without using
derivatives. With infinite spin singlet and spin triplet $NN$ scattering
lengths the four point functions are identical to Eq.~(\ref{A41}) at leading
order, and are therefore invariant under the Schr\"{o}dinger group. For
nucleons, the six point point functions can involve states with total spin 1/2
and 3/2. In the spin 1/2 channel a three body contact interaction with no
derivatives exists and is needed to renormalize the integral equation for three
body scattering\cite{birad}. This three body contact operator is expected to
introduce a new scale and therefore break scale and conformal invariance. In
the spin 3/2 channel \cite{bira32}, no three body operator is needed and this
amplitude is expected to respect the constraints from scale and conformal
invariance. Explicit verification of this would be interesting.

In this letter we derived Ward identities for amputated momentum space Green's
functions that follow from invariance under the Schr\"{o}dinger group. We also
examined implications of these constraints for 2-to-2 and 3-to-3 on-shell
scattering amplitudes. Motivated by recent developments in nuclear theory, we
considered a non-relativistic theory in the limit of infinite scattering length
and found it gives rise to a four point function which satisfies the Ward
identities which follow from Schr\"{o}dinger invariance.

We would like to thank John Preskill for a conversation which led to this paper,
and Jonathan Engel for a useful comment. This work was supported in part by the
Department of Energy under grant numbers DE-FG03-92-ER 40701 and
DOE-FG03-97ER40546. T.M. was also supported by a John A. McCone Fellowship.

 {\tighten

} 

\end{document}